# Asteroid Impact Effects And Their Immediate Hazards For Human Populations

**Authors**: Clemens M. Rumpf[1,*], Hugh G. Lewis[1], Peter M. Atkinson[2,3,4]

**Affiliations:**
[1] University of Southampton, Engineering and the Environment, Southampton, UK
[2] Lancaster University, Faculty of Science and Technology , Lancaster, UK
[3] University of Southampton, Geography and Environment, Southampton, UK
[4] Queen's University Belfast, School of Geography, Archaeology and Palaeoecology, Belfast, UK

[*]Corresponding Author. Clemens M. Rumpf (C.Rumpf@soton.ac.uk)

**Key Points:**

- Dominance of impact effects that are generated by asteroid impacts for every impactor diameter in the range of 15-400 m.

- Average casualty count estimation for impactors in the diameter range 0-400 m.

- Impactors over land are an order of magnitude more harmful than over water despite the generation of tsunamis.

## Abstract
A set of 50,000 artificial Earth impacting asteroids was used to obtain, for the first time, information about the dominance of individual impact effects such as wind blast, overpressure shock, thermal radiation, cratering, seismic shaking, ejecta deposition and tsunami for the loss of human life during an impact event for impactor sizes between 15 to 400 m and how the dominance of impact effects changes over size. Information about the dominance of each impact effect can enable disaster managers to plan for the most relevant effects in the event of an asteroid impact. Furthermore, the analysis of average casualty numbers per impactor shows that there is a significant difference in expected loss for airburst and surface impacts and that the average impact over land is an order of magnitude more dangerous than one over water.

## One Sentence Summary
Effect dominance varies over asteroid size and aerothermal effects are most harmful while impactors over land are more dangerous than over water.

## 1 Introduction
What are the consequences of an asteroid impact for the human population? This question is a significant driver for today's research activities that address the threat of asteroids that collide with the Earth [*Ailor et al.*, 2013]. Asteroid impacts produce an array of impact effects that can harm human populations. A list of seven such impact effects is recognized and described in [*Hills and Goda*, 1993; *Collins et al.*, 2005]. They are: wind blast, overpressure shock, thermal radiation, cratering, seismic shaking, ejecta deposition, and tsunami. The present work quantifies the contributions of each of these effects to overall losses due to an asteroid impact of a given size in a global setting.

Considerable work is available in the literature which addresses overall casualty numbers of asteroid impacts [*Stokes et al.*, 2003; *Harris*, 2008; *Shapiro et al.*, 2010; *Boslough*, 2013a; *Reinhardt et al.*, 2016]. Previous work has compared the loss of human life for impactors over land and water masses [*Stokes et al.*, 2003; *Shapiro et al.*, 2010] and these studies are currently being updated with an increased focus on individual impact effects [*Mathias et al.*, 2017; *Register et al.*, 2017]. Additional work has focused on the loss quantification of single impact effects such as tsunamis [*Chesley and Ward*, 2006] facilitating limited insight into the quantification of relative impact effect dominance. The focus of the present work is comparing the contribution (dominance) of the seven impact effects to overall loss and thereby providing a nuanced view of impact effect dominance.

To estimate loss of human life due to an asteroid impact, the severity of each impact effect needs to be calculated based on input parameters such as impactor size, impactor density, impact speed and impact angle. A suite of analytical impact effect models is provided in [*Collins et al.*, 2005] and it enables estimation of impact effect severity as a function of distance from the impact site (except for tsunamis). The literature provides examples for numerical codes that typically model few effects each in great detail [*Boslough and Crawford*, 2008; *Wünnemann et al.*, 2010; *Gisler et al.*, 2011; *Collins et al.*, 2012]. However, the high impactor count simulations performed here prohibited the use of numerically intensive codes. A suitable tsunami propagation model is presented in [*Rumpf et al.*, 2017] which utilizes ray tracing to determine affected coastlines on the global map depending on the impact location and calculates local coastal inundation based on bathymetry as well as topography data [*Patterson and US National Park Service*, 2015].

Here, the impact effects were propagated away from the impact location and across the local population utilizing global population data on a 2.5`x2.5` grid from 2015 [*CIESIN et al.*, 2005] to determine the number of affected people. The vulnerability of the affected population declines with increasing distance from the impact site as effect severity attenuates with distance. The vulnerability models used to determine local mortality, and, thus, overall casualties, are described in [*Rumpf et al.*, 2017]. Instead of propagating impact effects directly, a radius of destruction for each impact effect was estimated in [*Stokes et al.*, 2003; *Shapiro et al.*, 2010] based on work in [*Hills and Goda*, 1993]. Because global averages were of interest, simplifications regarding the population distribution were used in [*Chesley and Ward*, 2006] and [*Stokes et al.*, 2003; *Shapiro et al.*, 2010] by relying on statistical population numbers in coastal areas and by using the average land population density, respectively.

A large sample of artificial impactors was used in conjunction with the "Asteroid Risk Mitigation Optimization and Research" (ARMOR) tool [*Rumpf et al.*, 2016a, 2016b, 2017] to estimate the dominance of each impact effect and to produce results about the total loss potential of impactors in a global impact scenario as well as in impact scenarios over land and water masses. The impact scenarios covered the possible variations of impact speed and impact angle (see SI1.1).

## 2 THE IMPACTOR SAMPLE

Based on the distributions for impact location, speed and angle derived in the supplementary materials, an artificial impactor sample covering the globe and counting 50,000 impactors was randomly generated. To illustrate the spatial impact density, Figure 1 shows the sample's impact locations over Europe and the colour coding indicates randomly assigned impact angles. The method yielded 35,984 impactors, or 71.97% of the sample, that descended over water mirroring that 71% of the Earth's surface is covered with water. The sample of impactors was used to assess the dominance of individual impact effects for the population of Earth.

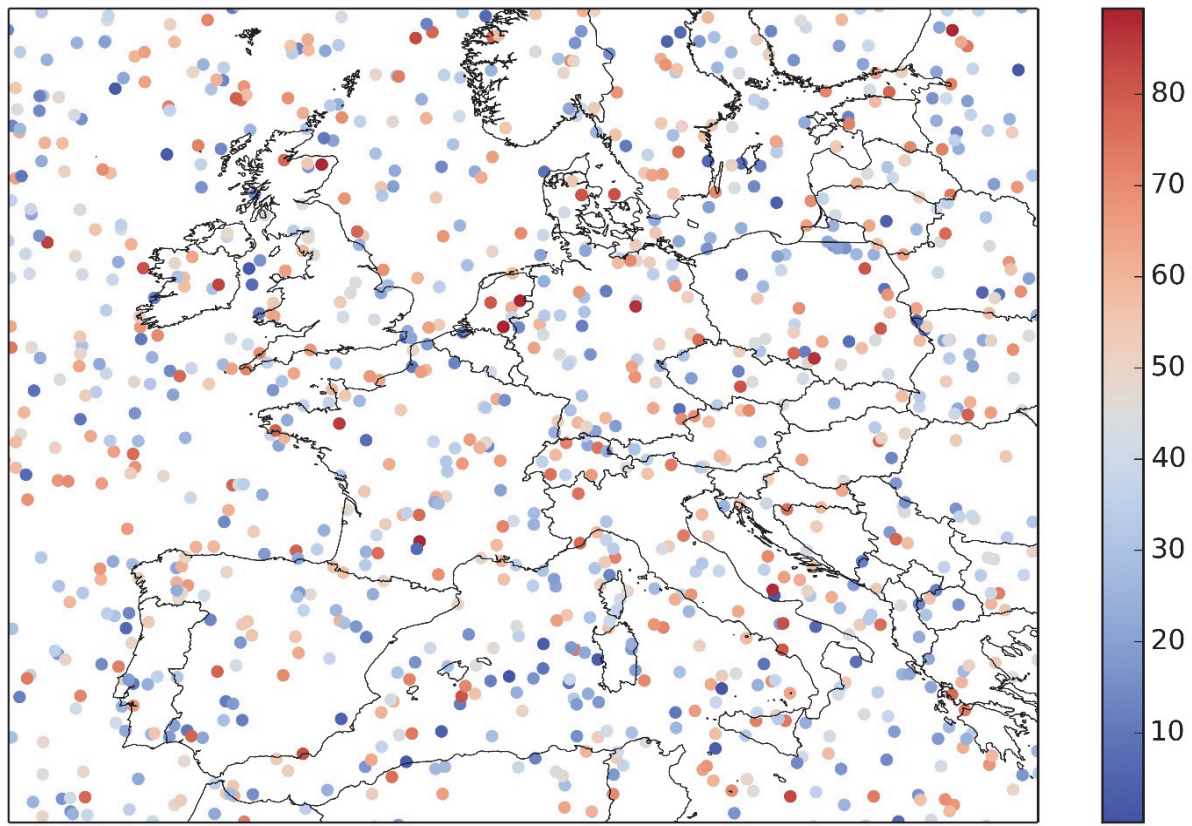

**Figure 1: Spatial visualisation of the realised set of impact locations over Western Europe. The colour of the markers reflects the impact angle in degrees where 90° is a vertical impact.**

## 3 FINDINGS

The dominance of asteroid impact effects was calculated, first, for a global impact scenario and, subsequently, for impacts over land and water masses separately. In the following figures, the total number of casualties recorded in each simulation run was divided by the sample size to obtain the average number of casualties per impactor.

The asteroid population exhibits a range of densities between 1000-8000 kg m$^{-3}$, however, about 80% of asteroids have a density between 1500-3500 kg m$^{-3}$ [*Zellner*, 1979; *Britt*, 2014; *Hanus et al.*, 2016]. Asteroid density can influence impact consequences significantly [*Hills and Goda*, 1993] and an asteroid density of 3100 kg m$^{-3}$ was assigned to the sample. The results are, thus, representative for this density value and provide a benchmark when considering density variations.

The results presented in Figure 2 show that asteroids of the assigned density, that reached the ground, were at least 56 m in diameter. All asteroids in the sample which were smaller than this size threshold experienced an airburst. While the combination of impact angle and speed has to be very specific to produce a surface impact at the threshold size, larger asteroids increasingly reached the surface because their bigger size allowed them to pass the atmosphere before disintegrating for a wider range of angle/speed combinations [*Toon and Covey*, 1997; *Collins et al.*, 2016]. The influence of density on this finding is such that, an increase in density will increase the chance of surface impacts, while a lower density will reduce that chance.

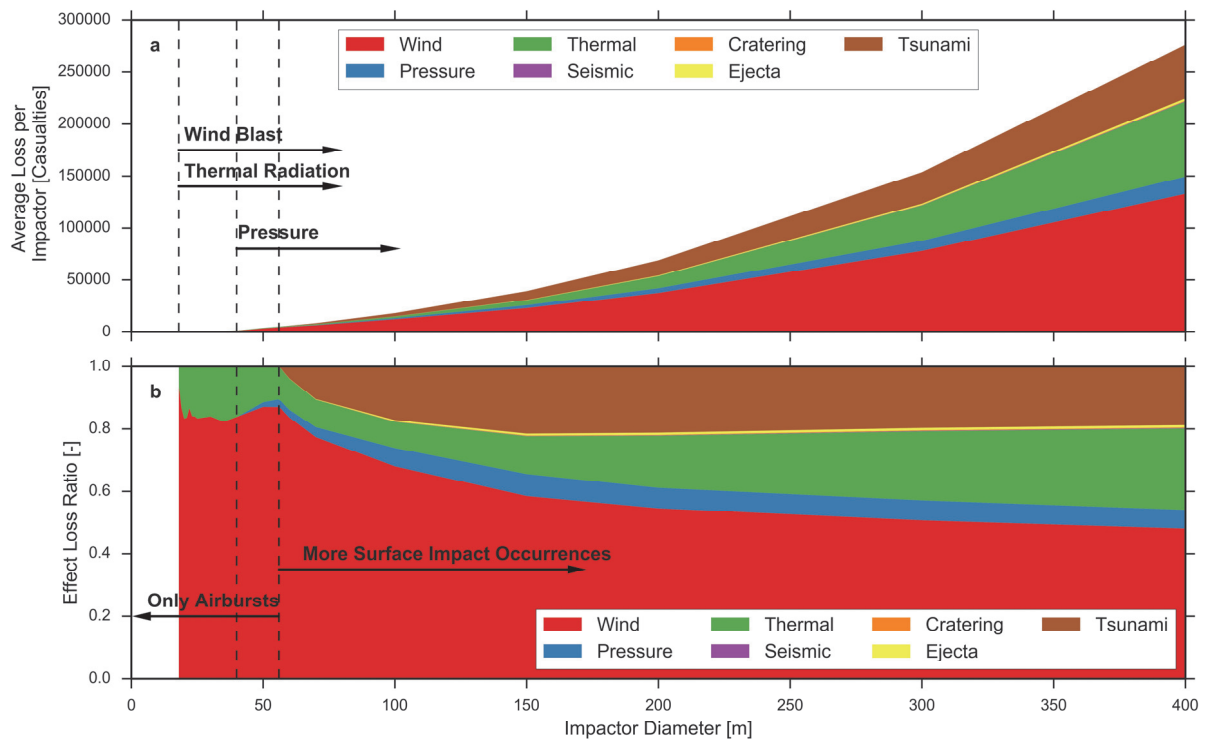

**Figure 2: Plot a shows the increase in average casualties per impactor size and highlights the increasing contribution by each impact effect. First casualties due to wind blast and thermal radiation occurred at 18 m. Impactors of 40 m produced the first pressure losses and first surface impacts were recorded for impactors larger than 56 m. Plot b shows the impact effect dominance distribution over the asteroid size range up to 400 m.**

For the chosen density, the minimum asteroid size to cause casualties was 18 m due to wind blast and thermal radiation. The harmful effect of an overpressure shock only became lethal for 40 m impactors (Figure 2). These findings correlate with observations made after the Chelyabinsk bolide event in 2013 where a 17-20 m object, travelling at 19 km s$^{-1}$ disintegrated mid-air [*Borovicka et al.*, 2013; *Brown et al.*, 2013]. Most of the damage and injuries during that event, were caused by the aerodynamic shock that knocked people to the ground and damaged structures and windows causing indirect injuries by flying glass shards. The population also reported burns, heat sensation and temporary blindness due to the intense electromagnetic radiation emitted by the meteor [*Popova et al.*, 2013]. The Chelyabinsk meteoroid was a shallow impactor that entered the atmosphere with an angle of 18° resulting in an airburst at an altitude of between 30 and 40 km [*Borovicka et al.*, 2013], which is consistent with the impact effect models used in this research that predicted an airburst altitude of 33 km [*Collins et al.*, 2005]. Given the possible impact conditions in terms of impact speed and angle distributions (Figure SI1c&d), these parameters reflect a medium energy event for an asteroid of this size, because of the shallow impact angle and no casualties were reported for the Chelyabinsk event. However, over 1,000 persons were injured [*Popova et al.*, 2013] and it is possible that an impactor of the same size with higher impact speed or steeper impact angle would have resulted in some casualties due to aerothermal effects. The Tunguska airburst in 1908 is another event for which considerable aerothermal damage was reported for a roughly 30-40 m sized object [*Boslough and Crawford*, 2008; *Artemieva and Shuvalov*, 2016]. During that event, over 2,000 km$^2$ of forest were flattened and trees in an area of 300 km$^2$ were burned by thermal radiation [*Nemtchinov et al.*, 1994; *Boslough and Crawford*, 2008]. While no human casualties have been reported for that event due to the remoteness of the impact location in Siberia, the released energy would certainly have sufficed to cause casualties in populated regions.

These observations are in line with land impact simulation results presented in Figure 3a and Figure 4a where aerothermal impact effects are predicted to cause significant loss.

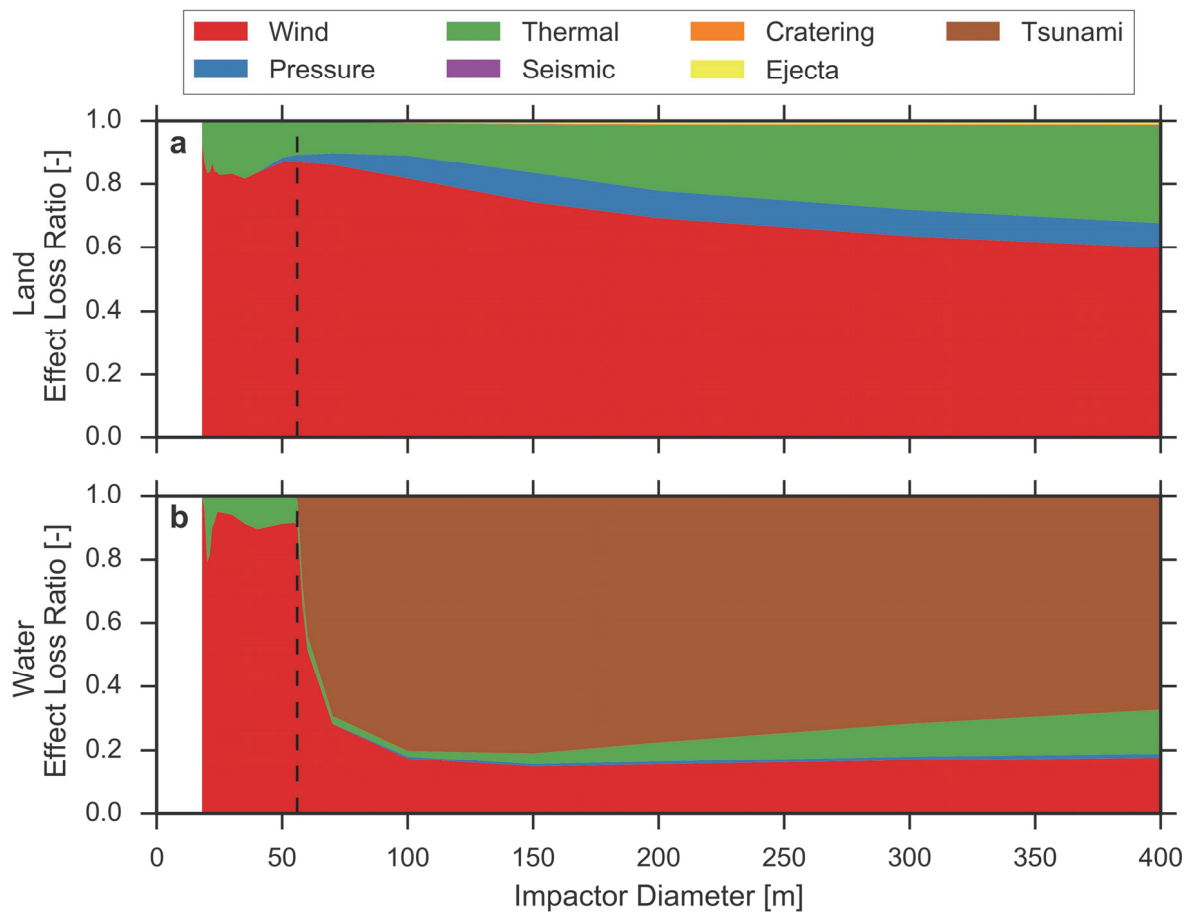

**Figure 3: Plot a visualizes the effect loss ratios for land impactors of a given size up to 400 m. Converesely, plot b shows these ratios for average water impactors of a given size. The vertical dashed line indicates the occurrence of first surface imapcts.**

The evolution of total average loss per impactor is visualized in Figure 4a on a semi-logarithmic scale for the global (red, middle line), land (green, upper line) and water (blue, lower line) impact scenario. The average land impactor is about one order of magnitude more dangerous than the average water impactor and this observation is supported by similar results in the updates to the reports [*Stokes et al.*, 2003; *Shapiro et al.*, 2010] [*Harris*, 2017]. Loss growth changes behaviour around the point of first surface impact occurrence. The average loss for impactor up to 50 m in diameter as a function of impactor size can be approximated by the fit (Pearson coefficient of 0.90):

$$y = 0.0835 \times 1.139^{1.748x} \qquad (1)$$

Similarly, the average loss for impactors which may reach the surface (>50 m), can be approximated (Pearson coefficient of 0.97) as a function of asteroid size with:

$$y = 4491.331 \times 1.0116^{0.984x} \qquad (2)$$

To gain insight into the variability of these results, best and worst case scenarios were designed intended to capture $\pm 1\sigma$ standard deviation [*Rumpf et al.*, 2017] and the results are expressed in Figure 4b as the ratio of the average global impact loss. The sensitivity analysis shows that results for small asteroid diameters may vary by a factor of two while larger asteroids show less sensitivity exhibiting variation of about +45/−30 % and these variations are in line with expected variations in previous work [*Stokes et al.*, 2003]. Figure 4c indicates the percentage of the impactor sample that contributed to loss generation. Values smaller than 50% correspond to a median impactor loss of 0.

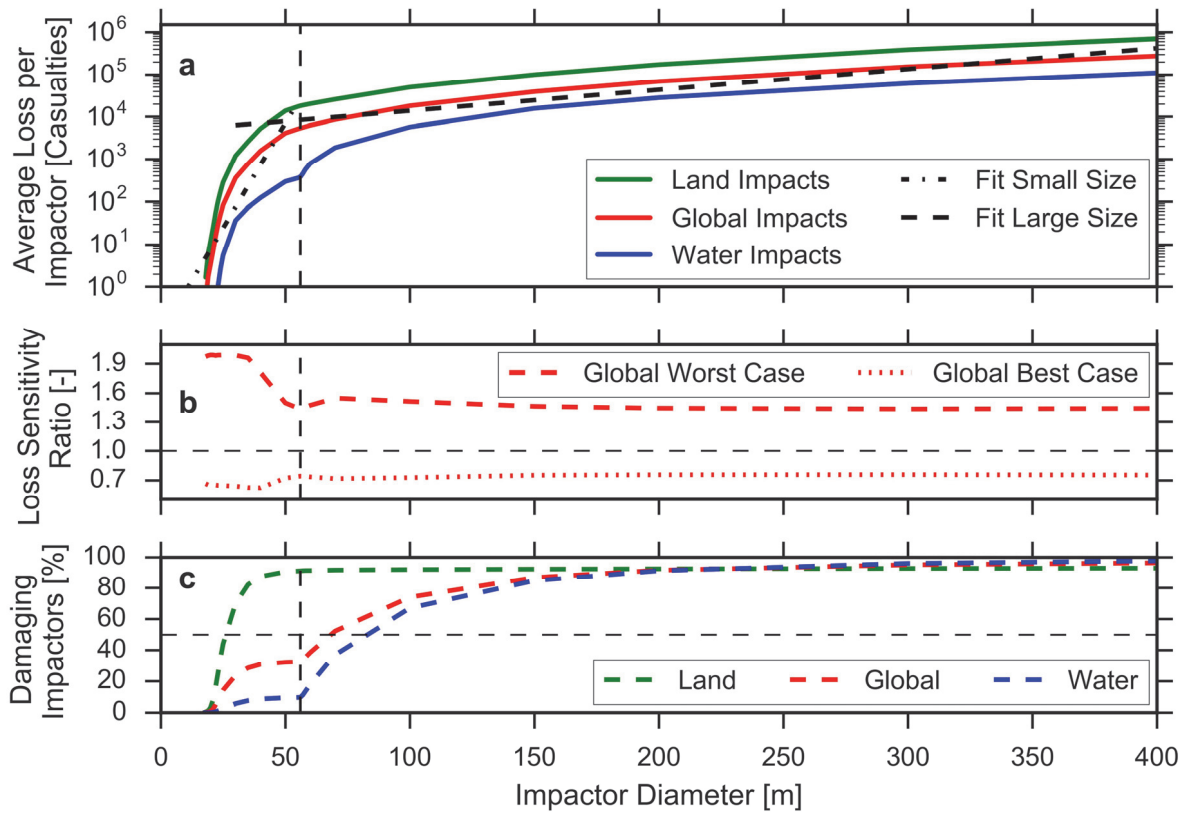

**Figure 4: Plot a presents average loss in the global, land and water impact scenario along with exponential fits for global airburst losses and losses due to larger impactors. Plot b indicates the variability in global loss numbers through correction factors for best/worst case scenarios. The expected case (factor 1) is marked with a horizontal dashed line. Plot c presents the percentage of impactors that contributed to loss generation in land, water or global scenarios. The 50% threshold is marked with a horizontal dashed line. To facilitate orientation, all plots show the size where first surface impacts occur with a vertical dashed line.**

The average loss per impactor increased exponentially with increasing impactor size and this is reflected in Figure 2a. Interestingly, the slope of the average loss function is larger in the airburst regime as shown by the fitted exponential functions (Equations 1 and 2) and in Figure 4a. This is partially owed to the fact that an increasing number of impactors harm the population (Figure 4c) but, in addition, aerothermal effects appeared to be more efficient at transforming their energy into loss. The latter statement is supported by the observation that a land impactor was an order of magnitude more harmful than a water impactor (Figure 4a). Furthermore, airbursts transform all available energy in aerothermal effects and do not split their energy for less harmful ground effects (Figure 2a). They are, thus, more efficient at depositing their destructive energy than larger impactors and this has relevance in connection with the fact that the asteroid catalogue is least complete (<1% discovered) in the small asteroid diameter range [*Harris and D'Abramo*, 2015]. Notably, the loss per impactor results agree well with previous work but extend the possibility for fatalities to smaller impactor sizes [*Chapman and Morrison*, 1994; *Stokes et al.*, 2003]. The residual risk from undiscovered asteroids might have to be corrected to smaller asteroid diameters [*Chapman and Morrison*, 1994; *Stokes et al.*, 2003; *Boslough*, 2013a; *Harris*, 2013].

Knowledge about the average number of total casualties per impactor can aid early decision-making about whether to deflect an asteroid or to evacuate the impact area when a new impactor is discovered and the corresponding sensitivity analysis provides insight into the expected spread in the casualty estimate. These results may be used in the future to facilitate a new asteroid hazard scale [*Binzel*, 2000; *Chesley et al.*, 2002; *Boslough*, 2013b; *Boslough et al.*, 2015].

While airbursting impactors appeared very efficient at depositing their energy, it is also important to understand that the loss outcome for individual impactors showed higher variation for small impactors. This is due to the shorter range, but high severity, of airburst effects compared to tsunamis. Figure 4c presents the percentage of the impactor sample that produced losses. In the global scenario, the median impactor loss was actually zero (corresponding to < 50% damaging impactors) for asteroids smaller than 60 m. This is owed to the fact that twice as many asteroids impact over water compared to land and the short range airburst effects do not reach any population. Focusing on the water impact scenario illustrates this point as only a small impactor fraction corresponding to near-coastal airbursts contributed to loss (Figure 4c). The sharp increase in loss for small impactor sizes can, thus, mainly be attributed to land impactors which are naturally close to populations. However, even for land impactors, the median impactor (smaller than 25 m) produced zero casualties illustrating that the average loss is driven by those impact events that hit close to densely populated areas and cause severe losses. In fact, the most damaging impactor was about four orders of magnitude more severe than the average loss for small impactors and this discrepancy decreases to two orders of magnitude at 400 m.

Land impact effect dominance is visualized in Figure 3a and these results show that wind blast in conjunction with overpressure shock are the most critical impact effects (since they act in concert) accounting for more than 60% of the losses up to 400 m. Wind blast and overpressure shock are generally treated in conjunction as they occur together [*Hills and Goda*, 1993]. They are presented separately in this work because their immediate harming mechanism on humans differ. Overpressure can rupture internal organs while a wind blast dislocates bodies and objects to cause harm [*Glasstone and Dolan*, 1977; *Rumpf et al.*, 2017]. Thermal radiation is significant but accounts for less than 30% losses. Notable is the increase in thermal radiation dominance for larger impact effects and this phenomenon is also present in the water impact scenario shown in Figure 3b. Not surprisingly, the most dominant effect for water impacts are tsunamis accounting for 70-80% losses depending on size. Together, land and water impacts make up the global scenario (Figure 2) with a correspondingly heavier weighing for the more dangerous land impactors (Figure 4a). The global scenario illustrates that aerodynamic effects dominate for all sizes (>50%). Thermal radiation is a significant concern and appears to increase in severity for larger impactors. Tsunamis have been a major concern in the planetary defense community but the results here suggest that they only contribute 20% to the overall threat of impacting asteroids.

Aerothermal effects dominate because they are caused by every impactor, while tsunamis can only be the result of an ocean impact. Furthermore, aerothermal losses are mainly caused by impactors over land which are naturally closer to population centers. In contrast, tsunamis can only reach near-coastal populations close to the coast because their inland reach is limited to a few kilometres. While the reach of tsunamis is far, these long propagation distances attenuate wave height significantly reducing population vulnerability during landfall. Furthermore, the initial wave height is limited by sea depth at the impact point [*Wünnemann et al.*, 2010]. The continental shelf forms a protective region [*Rumpf et al.*, 2017] around most coastlines reaching only about 100 m – 200 m depth and typically extending 65 km offshore [*The Editors of Encyclopædia Britannica*, 2016]. Even deep sea impacts of large asteroids are constricted by this upper boundary for wave height, while aerothermal effects can scale freely with impactor size and, thus, energy. In summary, it appears plausible that tsunamis contribute less than might be intuitively expected to global asteroid impact loss.

The findings provide valuable insight into which impact effects are most significant informing disaster managers about which effects the population should be prepared for in case of an impact. In the case of small impactors, aero-thermal effects are of greatest concern, and here, the population could seek shelter in a safe place such as a basement. For larger impactors, a complete evacuation might be necessary as high impact effect severity renders any affected region unsafe. For larger water impactors, tsunamis become a concern for near coastal populations which might need to be evacuated.

Conversely, knowledge about which impact effects are less significant is similarly valuable as it can help save resources otherwise spent on less critical impact effects. The influence of ejecta deposition is barely visible at the top of Figure 3a with a maximum contribution of 0.91%. Even less significant are the contributions of cratering and seismic shaking with a maximum of 0.2% and 0.17%, respectively. The results indicate that ground impact effects, such as cratering, seismic shaking and ejecta deposition, play a minor role in loss generation compared to other effects.

## 4 CONCLUSIONS

The analysis covered a wide range of possible impact conditions in terms of impact speed, angle and size using an impactor density of 3100 kg m$^{-3}$. Evaluation of this parameter space showed that the minimum asteroid size to cause fatalities was 18 m and that first surface impacts occur for asteroids with a minimum size of 56 m.

The total casualty estimation per impactor as a function of asteroid size was approximated by two exponential functions and these functions revealed that the loss generating mechanisms showed a significant change in behaviour around the surface impact size threshold. For smaller asteroids, only airbursts occurred and they appeared to be more efficient in transforming kinetic energy into loss than surface impacts. This finding may have implications for the assessment of residual asteroid impact risk of the yet undiscovered asteroid population which is biased towards smaller asteroid sizes.

Using the exponential description for total casualty estimation allows quick assessment of the possible threat when a new, impacting asteroid is discovered. Total casualty estimation also revealed that the average land impactor is about an order of magnitude more dangerous than the average water impactor. Aerothermal effects, dominated loss generation in the global setting. Equally importantly, the results provide evidence that effects such as cratering, seismic shaking and ejecta deposition provide only a minor contribution to overall loss. Tsunamis were the most significant effect for water impacts, but were less important globally. In summary, the results help to better understand the asteroid impact hazard, including which impact effects are most and least relevant, and can be of help in formulating an adequate response to the threat.

The small contribution of tsunamis to global loss was surprising but can be explained by initial wave height restriction due to the sea depth and wave height attenuation over distance whereas the other effects can scale freely with increasing impact energy and are naturally closer to populations. The data show, for the first time, how the dominance of impact effects changes for increasing impactor size.

## 5 Acknowledgement

The work is supported by the Marie Curie Initial Training Network Stardust, FP7-PEOPLE-2012-ITN, Grant Agreement 317185. The authors acknowledge the use of the IRIDIS High Performance Computing Facility at the University of Southampton and thank Dr Steven Chesley who provided the initial impactor sample. The authors would like to thank Alan W. Harris for reviewing this publication along with an anonymous reviewer.

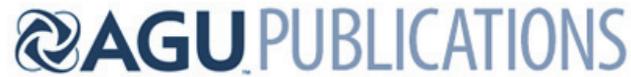

*Geophysical Research Letters*

Supporting Information for

# ASTEROID IMPACT EFFECTS AND THEIR IMMEDIATE HAZARDS FOR HUMAN POPULATIONS

Authors: Clemens M. Rumpf[1,*], Hugh G. Lewis[1], Peter M. Atkinson[2,3,4]

Affiliations:
[1] University of Southampton, Engineering and the Environment, Southampton, UK
[2] Lancaster University, Faculty of Science and Technology, Lancaster, UK
[3] University of Southampton, Geography and Environment, Southampton, UK
[4] Queen's University Belfast, School of Geography, Archaeology and Palaeoecology, Belfast, UK

*Corresponding Author. Clemens M. Rumpf (C.Rumpf@soton.ac.uk)

**Contents of this file**

    Text SI1 and SI2
    Figures SI1
    Supporting Information References

## *SI1 Impactor Sample*

To ensure that the analysis of impact effect dominance produces representative results, it was necessary to cover the Earth in a sufficiently high impactor density because the local impact environment, such as population count [*CIESIN et al.*, 2005] and surface type (land or water)[*Patterson and US National Park Service*, 2015], vary across the globe. Furthermore, the impactor sample needed to be representative in relation to the possible variations in impact angle and speed. In [*Chesley and Spahr*, 2004; *Grav et al.*, 2011], a random set of 10,006 artificial impactors was generated based on the solar system population of near-Earth objects (NEOs) and this set was used here to extract the distributions of impact location, angle and speed for Earth impactors. Subsequently, a fivefold larger artificial impactor sample of 50,000 impactors was generated randomly that reflects these distributions.

**SI1.1 Impact Characteristics**

The 10,006 strong impactor sample was assessed to determine the distributions for impact location (longitude and latitude) as well as impact angle and impact speed as shown in Figure SI1.

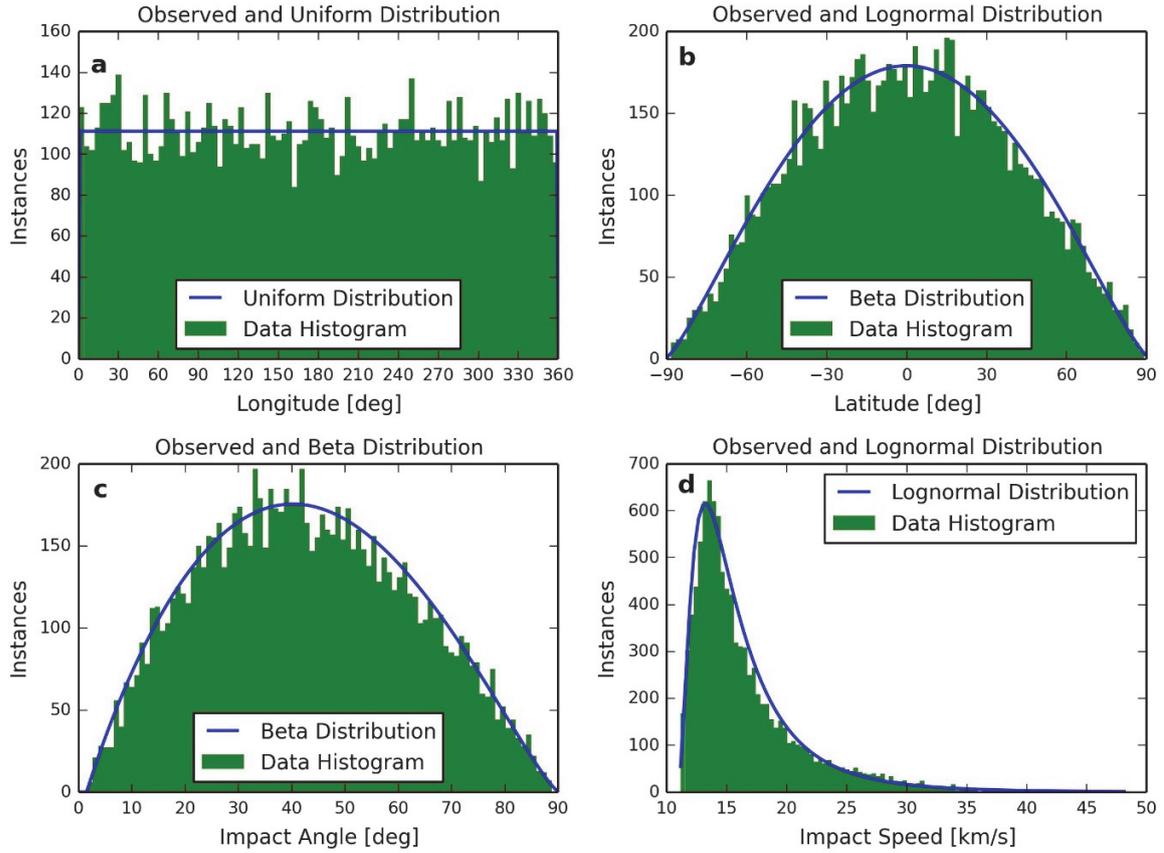

**Figure SI1: Distributions for impact location, angle and speed. Plot a shows the sample data (green bars) for impact location longitude and the corresponding uniform distribution fit (blue line). Plot b shows the beta distribution that was fitted to sample impact location latitude data. Plot c visualizes the sample data for impact angle and the fitted beta distribution. Plot d shows sample data for impact speed and the corresponding lognormal distribution fit.**

The longitudinal impact distribution was expected to be uniform [*Rumpf et al.*, 2016c] and the analysis of the impactor sample agreed with this expectation (Figure a) yielding a correlation coefficient of $R^2 = 0.9999$.

In terms of spatial impact density, the latitudinal impact density is expected to be nearly uniformly distributed as well [*Rumpf et al.*, 2016c]. However, in this analysis, the impact count (as opposed to spatial density) as a function of latitude was needed, and, because the area that corresponds to each latitude band decreases from the equator to the poles, the impact distribution is non-uniform with decreasing impactor count towards the poles. Figure b shows the best-fit result of a beta distribution to the observed latitudinal impact distribution with a correlation coefficient of $R^2 = 0.9999$. The corresponding probability density function is:

$$pdf_{beta}(x) = \frac{\Gamma(a+b)x^{(a-1)}(1-x)^{(b-1)}}{\Gamma(a)\Gamma(b)} \quad (1)$$

Where $a = 2.3188480$ and $b = 2.3293272$ and $\Gamma()$ is the Gamma function.

The impact angle (Figure c) is an important characteristic that determines the immediate environmental consequences in an asteroid impact event and a beta distribution was found to provide the best fit with the observed impact angle sample data ($R^2 = 0.9998$ for shape parameters $a = 1.967$, and $b = 2.248$ as applicable to Equation 1). It is noteworthy that the mean of this distribution is 42.87° (as opposed to the expected value of 45° as predicted by the analytical description of the problem [*Shoemaker et al.*, 1962]) showing that the distribution has a small positive skew favouring shallow impact angles. A possible reason for this finding is that Chesley and Spahr biased their population to reflect the impact frequencies of asteroids given their orbital speed [*Chesley and Spahr*, 2004]. Additional analysis was conducted to show that the impact angle is insensitive to impact latitude and no dependency could be detected.

Impact speed (Figure d) determines kinetic energy of an impact. The minimum impact speed of the sample was 11.138 km s$^{-1}$, and this lower bound can be attributed mainly to the Earth's escape velocity. The maximum observed impact speed was 48.119 km s$^{-1}$. It was found that a logarithmic normal distribution with probability density function:

$$pdf_{lognormal}(x) = \frac{1}{ax\sqrt{2\pi}} exp\left(-0.5\left(\frac{log(x)}{a}\right)^2\right) \qquad (2)$$

where $a = 0.7533689$, best fitted the observed distribution of impact speeds up to 50 km s$^{-1}$ which is the maximum allowed value in this analysis. For the impact speed range of 11.138 – 34.243 km s$^{-1}$, accounting for 98.63% of all impactors, this fit produced a correlation coefficient of $R^2 = 0.9903$. The mean of the lognormal distribution is 16.6 km s$^{-1}$. Additional analysis showed that impact speed has no appreciable dependency on impact latitude.

## *SI2 Assessing Impact Effect Dominance*

The impact simulations were conducted using the ARMOR tool [*Rumpf et al.*, 2016a, 2016b, 2016c] utilizing analytical impact effect models [*Collins et al.*, 2005] to calculate the environmental consequences of a given impact under consideration of impactor size, impact speed and angle, as well as location (i.e. local population and elevation data for tsunami propagation).

The impactors were assumed to be of spherical shape with a density of 3100 kg m$^{-3}$, reflecting the typical density of stony asteroids [*Zellner*, 1979; *Britt*, 2014; *Hanus et al.*, 2016]. Of special interest was how the dominance of each impact effect changes with increasing impactor size (diameter), effectively increasing the impactor's kinetic energy. Thus, a constant asteroid diameter was assigned to each simulation run starting with 17 m in the first simulation and increasing to 400 m in the last using a diameter resolution of 1 m in size regimes that lie around interesting phenomena, such as the occurrence of first casualties, or first surface impacts. During simulations, the total casualty numbers that were attributed to each impact effect were recorded to gain insight into which impact effects are most dominant in a specific size regime. Seven impact effects were considered and these are: wind blast, overpressure shock, thermal radiation, seismic shaking, cratering, ejecta deposition, and tsunami [*Collins et al.*, 2005; *Rumpf et al.*, 2016b]. The impact effects were propagated away from each impact site and vulnerability models (from [*Rumpf et al.*, 2016b], "expected" case) were employed to estimate the casualty count for each impact effect taking into account the population that lives in the affected area on a 4.6 km x 4.6 km global grid [*CIESIN et al.*, 2005].

The allocation of casualties to individual impact effects was accomplished by treating all impact effects independently of each other. In other words, each impact effect was allowed to interact with the *same* population. This can lead to a situation where the same person can become a casualty multiple times as different impact effects affect that same person. To mitigate double counting of casualties that would follow from this method, the simulations also calculated a total casualty count that followed the standard process of ARMOR's casualty estimation. In the standard process, casualty double counting is avoided and the population vulnerability is calculated taking into account all impact effects. Because population vulnerability to one impact effect ($V_{effect}$) is equivalent to the likelihood that any one person in that population dies, the chance that any one person survives an impact effect is simply $\lambda_{effect} = 1 - V_{effect}$, where $\lambda_{effect}$ is the likelihood of survival for that person. Considering that the person is affected by all impact effects in sequence, the combined chance of survival is the product of all effect survival probabilities:

$$\lambda_{combined} = \prod_{i=effects} \lambda_i \qquad (3)$$

Finally, the combined effect vulnerability is equivalent to the likelihood of that person not surviving all impact effects or:

$$V_{combined} = 1 - \lambda_{combined} \qquad (4)$$

The total casualty count determined with the standard vulnerability method was used to scale casualty counts of each impact effect such that the sum of the casualties of all effects was equal to the total casualty count.

**SI References**